\documentclass[10pt,pra,twocolumn,superscriptaddress,showpacs,floatfix]{revtex4}

\usepackage{graphicx}
\usepackage{amssymb}
\usepackage{times}
\usepackage{amsmath}
\usepackage{amsthm}
\usepackage{epstopdf}

\usepackage{hyperref}

\input epsf.tex
\usepackage{graphicx}
\usepackage{amsthm}
\usepackage{subfigure}

\begin{document}
\title{Security analysis of practical continuous-variable quantum key distribution systems under laser seeding attack}
\author{Yi Zheng}
\affiliation
 {College of Information Science and Technology, Northwest University, Xi'an 710127, Shaanxi, China
}
\author{Peng Huang}\thanks{Corresponding author: huang.peng@sjtu.edu.cn}
\affiliation
 {State Key Laboratory of Advanced Optical Communication Systems and Networks, and Center of Quantum Information Sensing and Processing, Shanghai Jiao Tong University, Shanghai 200240, China
}
\author{Anqi Huang}\thanks{Corresponding author: angelhuang.hn@gmail.com}
\affiliation
 {Institute for Quantum Information $\&$ State Key Laboratory of High Performance Computing, College of Computer, National University of Defense Technology, Changsha 410073, China
 }
\affiliation
 {Greatwall Quantum Laboratory, China Greatwall Technology, Changsha 410205, China
 }
\author{Jinye Peng}
\affiliation
{College of Information Science and Technology, Northwest University, Xi'an 710127, Shaanxi, China
}
\author{Guihua Zeng}
\affiliation
 {State Key Laboratory of Advanced Optical Communication Systems and Networks, and Center of Quantum Information Sensing and Processing, Shanghai Jiao Tong University, Shanghai 200240, China
}
\begin{abstract}
Here, we investigate the security of the practical one-way CVQKD and CV-MDI-QKD systems under laser seeding attack. In particular, Eve can inject a suitable light into the laser diodes of the light source modules in the two kinds of practical CVQKD systems, which results in the increased intensity of the generated optical signal. The parameter estimation under the attack shows that the secret key rates of these two schemes may be overestimated, which opens a security loophole for Eve to successfully perform an intercept-resend attack on these systems. To close this loophole, we propose a real-time monitoring scheme to precisely evaluate the secret key rates of these schemes. The analysis results indicate the implementation of the proposed monitoring scheme can effectively resist this potential attack.
\end{abstract}
\maketitle
\section{INTRODUCTION}\label{sec1}
Quantum key distribution (QKD) is a promising technology, which enables two authorized communication parties Alice and Bob to share a string of secret keys through an insecure quantum channel in the presence of a potential eavesdropper Eve \cite{Ekert1991Quantum,gisin2002quantum,weedbrook2012gaussian,grosshans2003quantum}. In theory, the basic laws of quantum physics guarantee the unconditional security of this technology \cite{Shor2000Simple,Lo1999Unconditional,leverrier2013security}. At present, QKD technology can be implemented by two kinds of different means, i.e., discrete-variable quantum key distribution (DVQKD) and continuous-variable quantum key distribution (CVQKD). Different from DVQKD systems, CVQKD systems rely on continuous modulation of the light field quadratures, which can be measured by utilizing the mature coherent detection technique instead of single-photon detection \cite{weedbrook2012gaussian,grosshans2003quantum}. Therefore, CVQKD systems can be well compatible with the classical optical communication systems. In particular, CVQKD with the Gaussian-modulated coherent states (GMCS) is one well-known protocol, which has been proven to be secure against the collective and coherent attacks \cite{leverrier2013security}. Over the past years, the GMCS CVQKD scheme has been experimentally implemented by many research groups in laboratories and in field environment \cite{qi2007experimental,fossier2009field,Jouguet2012Experimental,huang2016field}. In this work, we also focus on the investigation of the GMCS CVQKD schemes.

As we all known that the implemented devices (e.g., laser, modulators, detetors) of the GMCS CVQKD schemes is assumed to be secure and perfect in the security proofs \cite{leverrier2013security}. In fact, however, there are some direct or indirect imperfections in practical GMCS CVQKD systems \cite{Jouguet2012Analysis}. These imperfections can be divided into two categories. First, some imperfections may open several security loopholes. In particular, Eve can exploit these loopholes to steal key information without being detected, which seriously threatens the practical security of the systems. This is an effective quantum hacking strategy, such as the local oscillator (LO) fluctuation attack \cite{Ma2014Local}, the LO calibration attack \cite{Jouguet2013Preventing}, the wavelength attack \cite{Huang2013Quantum,Ma2014Wavelength,huang2014quantum}, the saturation attack \cite{Qin2016Quantum}, finite sampling bandwidth effects \cite{Wang2016Practical}, homodyne detector blinding attack \cite{qin2018homodyne}, jitter in clock synchronization \cite{xie2018practical}, and the polarization attack \cite{zhao2018polarization}. Second, the other imperfections can simply deteriorate the performance of the systems, such as imperfect phase compensation \cite{huang2015security}, finite-size effects \cite{leverrier2010finite}, and the noisy coherent states \cite{liu2017imperfect,filip2008continuous,usenko2010feasibility,shen2011continuous,yang2012source}. These imperfections hinder the commercial application of CVQKD.

Subsequently, several strategies have been designed to remove these imperfections. For the first imperfections that can open security loopholes, countermeasures are proposed to improve existing systems. For example, a real-time shot-noise measurement (RTSNM) scheme is used to resist the attacks originating from the local oscillator (LO) signal \cite{Jouguet2013Preventing,Liu2017}. Then, a local LO (LLO) CVQKD scheme is designed and implemented experimentally, which can fundamentally close the security loophole originates from LO \cite{soh2015self,Qi2015Generating,huang2015high-speed,wang2018pilot,Tao2018High}. Another attractive approach is to improve GMCS CVQKD protocol directly, i.e., continuous-variable measurement-device-independent quantum key distribution with the Gaussian-modulated coherent states (GMCS CV-MDI-QKD) protocol, which is immune to all detector side-channel attacks \cite{pirandola2015high-rate,ma2014gaussian,li2014continuous-variable,Zhang2017Finite,papanastasiou2017finite,ma2018continuous}. To remove the second imperfections, the reasonable noise models are needed to precisely evaluate the performance of the system. It is important to note that the above countermeasures do not close all potential loopholes, and new proposed attacks may defeat the CVQKD system. Therefore, the discoveries and preventions of the concealed security loopholes are vital to the commercial application of CVQKD.

Light source is one of key devices for the implementation of QKD systems, which is assumed to be trusted in previous research. For instance, in CVQKD, the noisy Gaussian source is well studied and modeled \cite{shen2011continuous,yang2012source}. However, the parameters of the light source may be actively tampered by Eve \cite{huang2019laser,sun2015effect,pang2019hacking}. In particular, based on the framework of GMCS CV-MDI-QKD, the sources become the final battlefield between the authorized communication parties and Eve. Therefore, the effects of the tampered source should be considered for the security analysis of practical CVQKD systems, which has not been well studied.

More recently, Huang $\it{et}$ $\it{al}$. proposed an efficient quantum hacking strategy related with light source to attack the DVQKD systems, which is called as laser seeding attack \cite{huang2019laser}. In this quantum hacking scheme, Eve can inject bright light into the laser diode of the systems to actively open a loophole. In this paper, inspired by this quantum hacking attack in DVQKD, we study the security of practical CVQKD systems under the laser seeding attack. Here, we focus on several well-known CVQKD protocols, i.e., the standard one-way GMCS CVQKD and GMCS CV-MDI-QKD schemes. More specifically, we first reveal that Eve can exploit the laser seeding attack to make the intensity of the transmitted Gaussian-modulated coherent states increased. Then, we find that the laser seeding attack makes the quantum channel excess noises of these two systems underestimated. Subsequently, we show that the secret key rates of these two systems are overestimated by Alice and Bob under the attack. These imperfect evaluative results are coincident with the security analysis results of CVQKD under the effects of the reduced optical attenuation caused by the laser damage attack \cite{zheng2019practical,huang2019damage,bugge2014laser}, which indicates that the laser seeding attack can also open a security loophole for Eve to perform an intercept-resend attack on these two kinds of CVQKD systems without being detected. In particular, although the CV-MDI-QKD protocols can remove all side channels originate from measurement unit, we observe that it is more vulnerable than the one-way CVQKD schemes to the laser seeding attack. Finally, we design a countermeasure to resist the laser seeding attack, where the intensity of optical signal generated by light source is monitored by the authorized communication parties in real time. The analysis result indicates that the legitimate communication parties can precisely evaluate the channel parameters to accurately calculate the secret key rate of these two systems through this scheme.

This paper is organized as follows. In Sec. \ref{sec2}, the laser seeding attack is described and modeled. Then, the security of various CVQKD systems under the laser seeding attack is studied in Sec. \ref{sec3}. In Sec. \ref{sec4}, we investigate the countermeasure to resist the laser seeding attack. Finally, conclusions are presented in Sec. \ref{sec5}.
\section{PRINCIPLE OF THE LASER SEEDING ATTACK}\label{sec2}
\subsection{Scheme of laser seeding attack}\label{sec21}
\begin{figure}[!h]\center
\centering
\resizebox{8.6cm}{!}{
\includegraphics{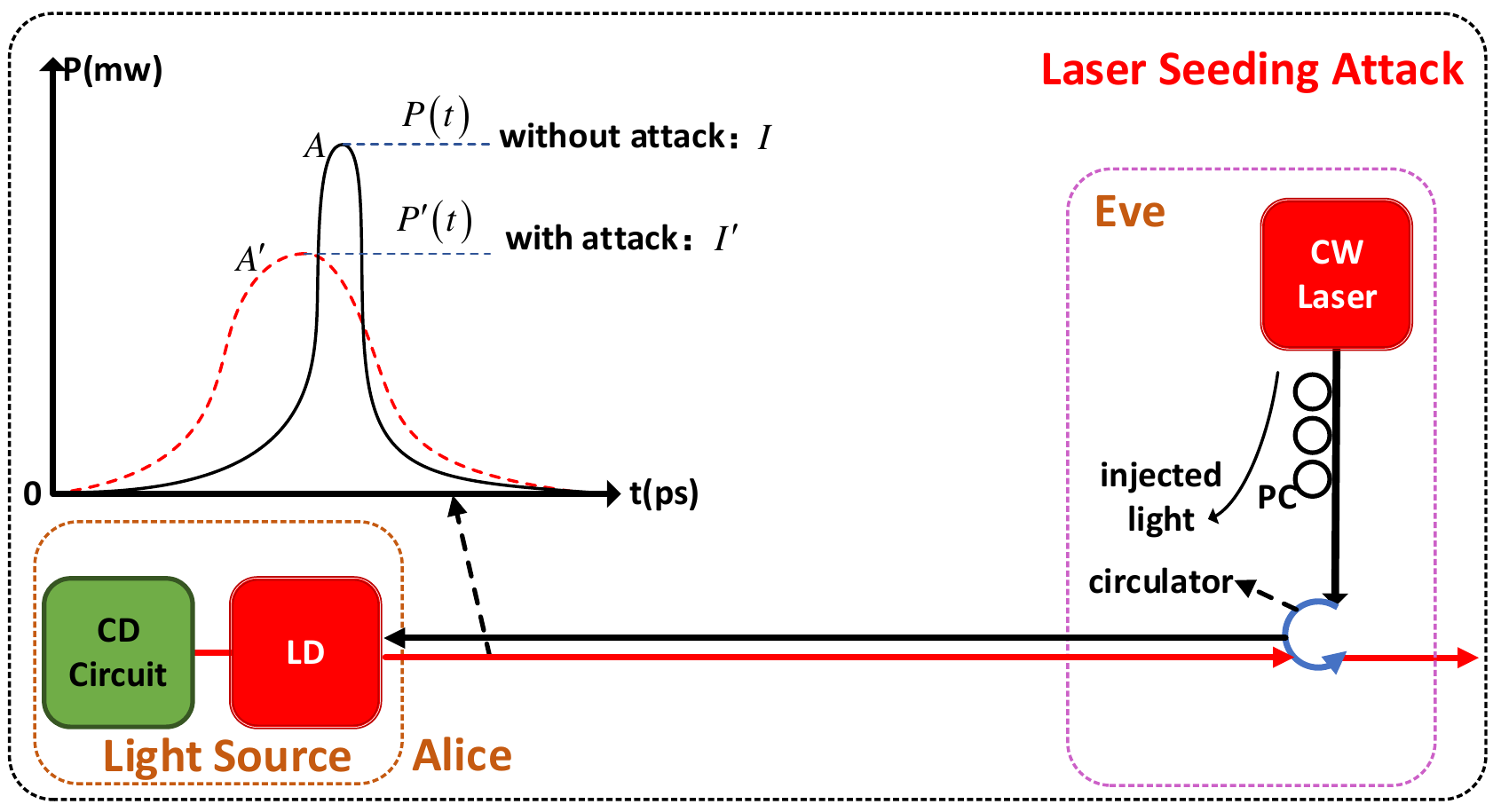}}
\caption{The schematic diagram of the laser seeding attack. CD Circuit, current driver circuit; CW Laser, continuous-wave laser; LD, laser diode; PC, polarization controller; $P(t)$, the power of the optical signal emitted by a laser diode without the laser seeding attack; $P^\prime(t)$, the power of the optical signal emitted by a laser diode with the attack; $A$, the generated signal without the attack; $A^\prime$, the generated signal with attack; $I$, the intensity of the pulse signal generated by the light source module without the attack; $I^\prime$, the intensity of the pulse signal generated by the light source module with the attack.}
\label{FIG1}
\end{figure}

In Ref. \cite{huang2019laser}, Huang $\it{et}$ $\it{al}$. proposed the laser seeding attack and demonstrated that Eve may perform the attack in the light source of a practical DVQKD system to steal key information without being detected, which seriously destroys the practical security of the system. Fig. \ref{FIG1} shows the scheme of the laser seeding attack clearly according to the experimental results in Ref. \cite{huang2019laser}. Specifically, Eve can utilize a tunable continuous-wave laser to inject a bright light with a proper wavelength into the semiconductor laser diode of a DVQKD system via quantum channel, where the semiconductor laser diode generates the optical signals driven by the electrical signals. In particular, a polarization controller is used for adjusting the polarization of the injected light signal to maximize the injection efficiency. According to the analyses in Ref. \cite{huang2019laser}, we describe the two curves of power of the optical signal generated by the laser diode varying with time in the laser seeding attack case and ideal case, respectively. The laser seeding attack will cause two main effects on the ideal curve, which are shown in Fig. \ref{FIG1}. The first impact is that the curve becomes wider with a much higher and longer tail. The other influence is that the peak of the curve shifts to earlier compared with the ideal situation. Here, we use $P(t)$ and $P^\prime(t)$ to represent the power of the output optical signal in ideal situation and with the attack, respectively. Then, the intensity of the optical signal prepared by light source without the laser seeding attack can be calculated as
\begin{equation}
\label{eq1}
I=\mu\int_0^T P(t) \mathrm{d}t,
\end{equation}
where $T$ is the period of the optical pulse emitted by laser diode, $\mu$ is a certain coefficient related with detection. Here, we assume that the parameters $T$ and $\mu$ are fixed and unaffected by the laser seeding attack. Correspondingly, the intensity $I^\prime$ of the optical signal generated by the attacked light source can also be acquired by Eq. (\ref{eq1}). It is obvious that the intensity of the optical signal prepared by the light source gets larger under the effects of the laser seeding attack. For simplicity, we assume that $I^\prime=gI(g>1)$ in the following analysis. Here, $g$ reflects the power of the laser seeding attack.

Similarly, in practical CVQKD systems, light source is also a key device, which can be used for generating information carrier signal and LO signal in the transmitter Alice. In particular, light source is also used in Bob's side for a CV-MDI-QKD system. In implementations of CVQKD, the semiconductor laser diode is also widely used to generate the optical signals driven by the electrical signals. For example, 100 ns coherent light pulses can be prepared for CVQKD by using a 1550 nm laser diode at a repetition rate of 1 Mhz \cite{Jouguet2012Experimental}. Therefore, Eve may perform the laser seeding attack in a CVQKD system. In the following sections, we will focus on the theoretical security research of various CVQKD systems under the laser seeding attack.
\subsection{The effects of the laser seeding attack in a CVQKD system}\label{sec22}
In the practical implementation of a standard one-way GMCS CVQKD system, Alice modulates the random key information to the pulse signal $A$, which can result in a series of Gaussian-modulated coherent states $|\alpha\rangle$ \cite{weedbrook2012gaussian}. After optical attenuation, we use $|\alpha_{A_0}\rangle$ to indicate the transmitted coherent states. Based on the phase space, $|\alpha_{A_0}\rangle$ can be written as
\begin{equation}
\label{eq2}
\begin{split}
&|\alpha_{A_0}\rangle=|\alpha_{A_0}|e^{i\theta}=x_{A_0}+ip_{A_0},\\
&x_{A_0}=|\alpha_{A_0}|\cos\theta,p_{A_0}=|\alpha_{A_0}|\sin\theta,
\end{split}
\end{equation}
where $|\alpha_{A_0}|$ and $\theta$ indicate the amplitude and phase of the transmitted Gaussian-modulated optical signal $A_0$, respectively. In particular, $x_{A_0}$ and $p_{A_0}$ are two independent quadratures variables with identical variance $V_{A_0}$ and zero mean \cite{grosshans2003quantum,weedbrook2012gaussian}.
\begin{figure}[!h]\center
\centering
\resizebox{6.5cm}{!}{
\includegraphics{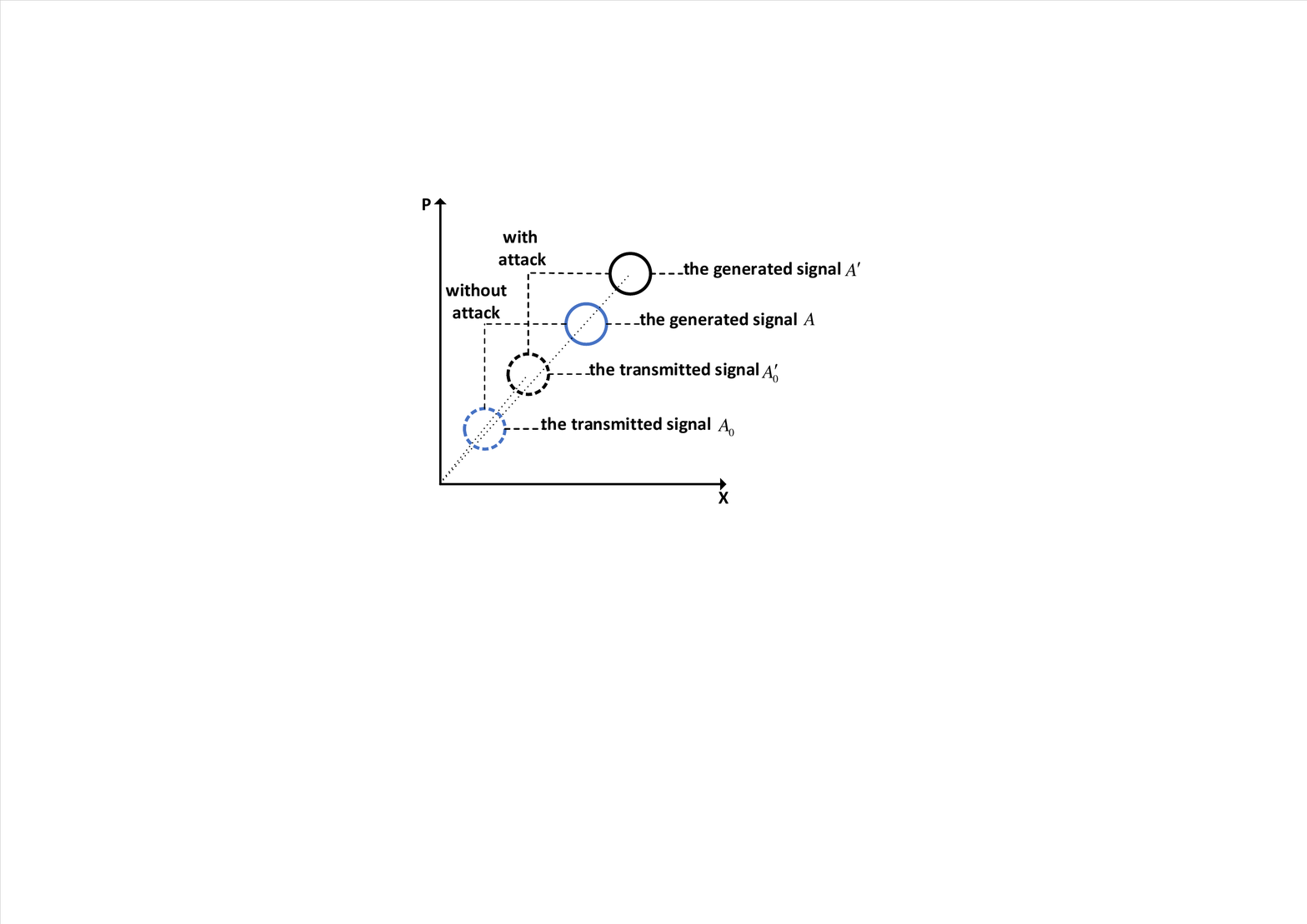}}
\caption{The expression of the transmitted Gaussian-modulated coherent states in the phase space under the lase seeding attack.}
\label{FIG2}
\end{figure}

However, the generated optical signal $A$ can change under the laser seeding attack. Therefore, $x_{A_0}$, $p_{A_0}$, and $V_{A_0}$ may deviate from their ideal values due to the influences of the attack, which is revealed in Fig. \ref{FIG2} based on the phase space. Since $I\propto|\alpha_{A_0}|$, the changes of $x_{A_0}$, $p_{A_0}$, and $V_{A_0}$ are as follows:
\begin{equation}
\label{eq3}
\begin{split}
&x^\prime_{A_0}=\sqrt{g}x_{A_0},p^\prime_{A_0}=\sqrt{g}p_{A_0},\\
&V^\prime_{A_0}=gV_{A_0},
\end{split}
\end{equation}
where $x^\prime_{A_0}$ and $p^\prime_{A_0}$ are two independent quadratures variables of the transmitted quantum signal $A^\prime_0$ with the attack, $V^\prime_{A_0}$ are the variance of $x^\prime_{A_0}$ or $p^\prime_{A_0}$. Similarly, the effects also exist in a CV-MDI-QKD system.

In addition, it is important to note that the intensity of LO signal can also become large with the increase of the intensity of the generated optical signal $A$. In a practical CVQKD system, there are many reasons for the increase of the intensity of LO signal, such as the decrease of optical attenuation \cite{zheng2019practical}. Although the origin of the change is ambiguous for Alice and Bob, the real-time shot-noise measurement technique can eliminate the impact. Therefore, we do not have to consider the influences of the increased intensity of the LO signal in the following analysis.
\section{SECURITY ANALYSIS}\label{sec3}
\subsection{Security of a one-way GMCS CVQKD system under the laser seeding attack}\label{sec31}
The analyses in Sec. \ref{sec2} indicate that the laser seeding attack will lead to the increased intensity of the transmitted Gaussian-modulated coherent states, which is the same as the influences of the reduced optical attenuation caused by the laser damage attack \cite{bugge2014laser,makarov2016creation,zheng2019practical,huang2018quantum,huang2019damage}. Therefore, based on the analyses in Ref. \cite{zheng2019practical}, it is feasible that Eve can unconsciously steal key information shared by Alice and Bob in a one-way GMCS CVQKD system by using the laser seeding attack. The analysis result indicates that the enhancement of the transmitted Gaussian-modulated coherent states can open a security loophole for Eve to attack a one-way GMCS CVQKD system without trace.
\begin{figure}[!h]\center
\centering
\resizebox{7.6cm}{!}{
\includegraphics{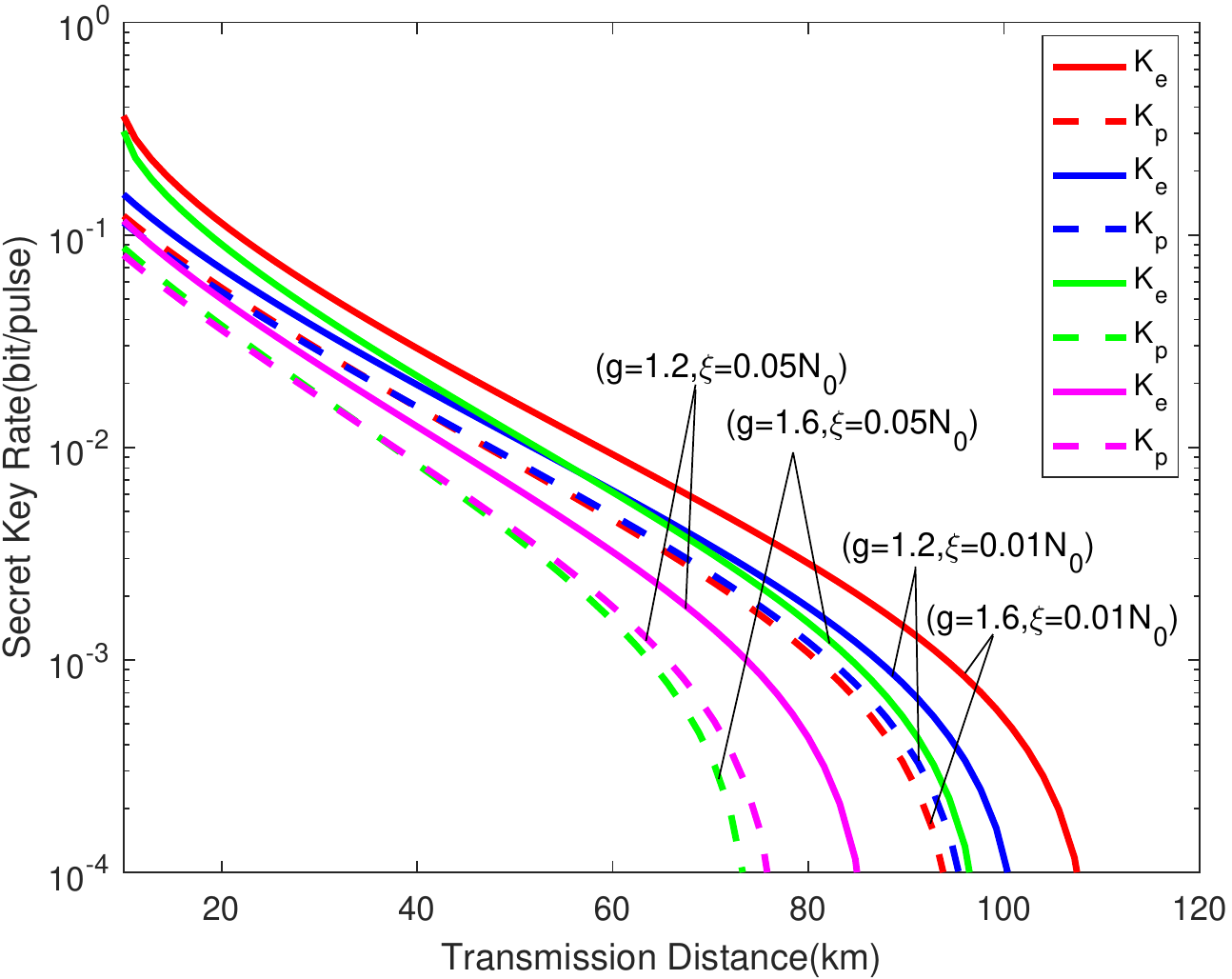}}
\caption{Secret key rate vs transmission distance for different powers $g$ of the laser seeding attack when $\varepsilon=0.01,0.05$, respectively. Solid curves from top to bottom represent the relations between the evaluated secret key rate $K_e$ and the transmission distance. Dotted curves show the corresponding practical secret key rate $K_p$ vs transmission distance under different situations. The fiber loss is 0.2 $\text{dB}/\text{km}$.}
\label{FIG3}
\end{figure}

Further, whatever the reason for the increased intensity of the transmitted Gaussian-modulated coherent states, we can use the analysis method presented in \cite{zheng2019practical} to investigate the practical security of the system under this loophole. Thus, we no longer describe the calculation process of the secret key rate for the system under the laser seeding attack. Fig. \ref{FIG3} depicts the relationship between the secret key rate and the transmission distance for the one-way GMCS CVQKD system under the laser seeding attack when $\varepsilon=0.01,0.05$. The fixed parameters for the simulation are set as: $V_{A_0}=4, \eta=0.5, \nu_{\text{el}}=0.01, \beta=95\%, \epsilon=10^{-10}, m=0.5\times N$, respectively.

It is clear that the evaluative secret key rate $K_e$ under the laser seeding attack are overestimated compared with the practical secret key rate $K_p$ in the same situation. These results also demonstrate that the laser seeding attack can open a security loophole for Eve to perform an intercept-resend attack in a practical one-way CVQKD system. In particular, the gap between the estimated secret key rate and the corresponding practical secret key rate represents the key information that can be acquired by Eve through the intercept-resend attack. We find that the leaking of the secret key information ascends with the power of the laser seeding attack. In addition, Eve can acquire more secret key information in the case of a larger excess noise $\varepsilon$ under the same attack power of laser seeding attack.
\subsection{Security of GMCS CV-MDI-QKD systems under the laser seeding attack}\label{sec32}
In GMCS CV-MDI-QKD systems, the two sources become the only region that can be exploited by Eve. Here, these sources are the same as the source of one-way GMCS CVQKD systems. Therefore, it is possible that Eve attacks these sources actively. In the following analysis, we will investigate the practical security of CV-MDI-QKD systems under the laser seeding attack in detail.
\subsubsection{The estimated channel parameters under the laser seeding attack}
\begin{figure}[!h]\center
\centering
\resizebox{8.5cm}{!}{
\includegraphics{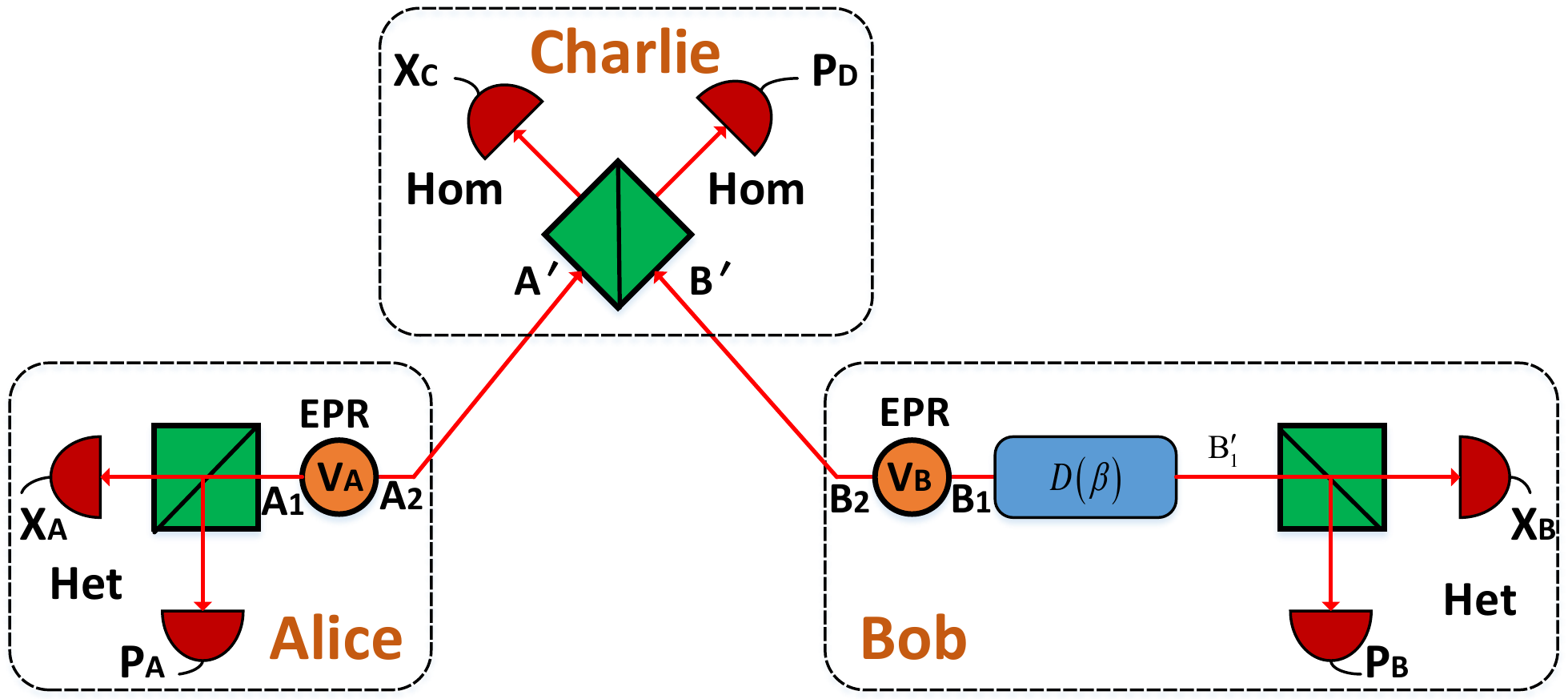}}
\caption{EB scheme of the CV-MDI-QKD protocols. EPR, two-mode squeezed state; Het, heterodyne detection; Hom, homodyne detection; $D\left( \beta  \right)$, displacement operation.}
\label{FIG4}
\end{figure}

Although CV-MDI-QKD can remove all known or unknown side-channel attacks on detectors, Eve may perform the laser seeding attack on the light source module in a practical CV-MDI-QKD system. Therefore, it is essential that the practical security of a CV-MDI-QKD system under the laser seeding attack is investigated. In Fig. \ref{FIG4}, we describe the equivalent entanglement-based (EB) model of the CV-MDI-QKD schemes. Specifically, Alice and Bob first generate one two-mode squeezed state with variance ${V_A} + 1$ and ${V_B}+ 1$, respectively. Here, the mode ${A_1}$ (${B_1}$) is retained by Alice (Bob), the other mode ${A_{\rm{2}}}$ (${B_{\rm{2}}}$) is sent to an untrusted third party Charlie through the quantum channel with length $L_{AC}$ ($L_{BC}$). The total transmission distance $L_{AB}$ is equal to ${L_{AC}} + {L_{BC}}$. Subsequently, Charlie interferes two modes $A'$ and $B'$ at a beam splitter ($\text{BS}$) with two output modes $C$ and $D$. Then, both the quadrature variable $x_C$ of mode $C$ and quadrature variable $p_D$ of mode $D$ are measured by Charlie through homodyne detection, and he announces the measurement results $\left\{ {{x_C},{p_D}} \right\}$ through a public channel \cite{pirandola2015high-rate,ma2014gaussian}. Finally, Bob modifies mode ${B_1}$ to ${B'_{\rm{1}}}$ by displacement operation $D\left( \beta  \right)$, where $\beta  = g_m\left( {{x_C} + i{p_D}} \right)$, and $g_m$ represents the gain of the displacement operation. After through these procedures, mode ${A_1}$ and ${B'_{\rm{1}}}$ become entangled. Accordingly, Alice and Bob will share a group correlated vectors $X = \left\{ {\left( {{{x_{A_0}}_i},{{x_{B_0}}_i}} \right)\left| {i = 1,2, \ldots ,N} \right.} \right\}$ or $P = \left\{ {\left( {{{p_{A_0}}_i},{{p_{B_0}}_i}} \right)\left| {i = 1,2, \ldots ,N} \right.} \right\}$ after the quadratures of mode ${B'_{\rm{1}}}$ and mod ${A_1}$ are measured by employing heterodyne detection. It is notable that Alice and Bob implement information reconciliation and privacy amplification to obtain a string of secret key.

It is obvious that there are two quantum channels which satisfy with linear model in a CV-MDI-QKD system. Here, the excess noise and transmittance of the quantum channel between Alice (Bob) and Charlie are expressed by ${\varepsilon _{AC}}$ (${\varepsilon _{BC}}$) and ${T_{AC}}$ (${T_{BC}}$). Since the transmission of the Gaussian-modulated quantum coherent states before the interference in Charlie's apparatus is the same with the one-way GMCS CVQKD systems, the involved two quantum channels can also be modeled as a normal linear model with the following relations:
\begin{equation}
\label{eq4}
\begin{split}
x_{A^\prime_0}&=t_{AC}x_{A_0}+z_{AC},\\
p_{A^\prime_0}&=t_{AC}p_{A_0}+z_{AC},\\
x_{B^\prime_0}&=t_{BC}x_{B_0}+z_{BC},\\
p_{B^\prime_0}&=t_{BC}p_{B_0}+z_{BC},
\end{split}
\end{equation}
where $x_{A_0}$ ($p_{A_0}$), $x_{A^\prime_0}$ ($p_{A^\prime_0}$), $x_{B_0}$ ($p_{B_0}$) and $x_{B^\prime_0}$ ($p_{B^\prime_0}$) are the values of the quadrature variables of mode $A_2$, $A^\prime$, $B_2$ and $B^\prime$, respectively. Here, $t_{AC}=\sqrt{T_{AC}}$, $t_{BC}=\sqrt{T_{BC}}$, $z_{AC}$ and $z_{BC}$ are the total noises in the two quantum channels that obeys two centered normal distribution with variance $\sigma^2_{AC}=T_{AC} \xi_{AC}+N_0$ and $\sigma^2_{BC}=T_{BC} \xi_{BC}+N_0$, respectively. These variances include shot noise $N_0$, channel excess noises $\xi_{AC}$ and $\xi_{BC}$.

Based on the above analyses, we can obtain the following relations \cite{ma2014gaussian,Zhang2017Finite,li2014continuous-variable,papanastasiou2017finite}:
\begin{equation}
\label{eq5}
\begin{split}
\langle x^2_{A_0} \rangle=&\langle p^2_{A_0}\rangle =V_{x_{A_0}}=V_{p_{A_0}},\\
\langle x^2_{B_0}\rangle=&\langle p^2_{B_0} \rangle=V_{x_{B_0}}=V_{p_{B_0}},\\
\langle x^2_C \rangle=&\langle p^2_D \rangle=\frac{1}{2}\eta(T_{AC}V_{x_{A_0}}+T_{BC}V_{x_{B_0}})+N_0\\
&+V_{el}+\frac{1}{2}\eta(T_{AC}\xi_{AC}+T_{BC}\xi_{BC}),\\
\langle x_{A_0}x_C\rangle=&\sqrt{\frac{\eta T_{AC}}{2}}V_{x_{A_0}},\langle p_{B_0}p_D\rangle=\sqrt{\frac{\eta T_{BC}}{2}}V_{p_{B_0}},\\
\langle x_Cx_D\rangle=&\langle p_Cp_D\rangle=\frac{1}{2}\eta(T_{AC}V_{x_{A_0}}-T_{BC}V_{x_{B_0}})\\
&+\frac{1}{2}\eta(T_{AC}\xi_{AC}-T_{BC}\xi_{BC}).
\end{split}
\end{equation}
According to Eq. (\ref{eq5}), we can further get
\begin{equation}
\label{eq6}
\begin{split}
\langle x^2_C\rangle+\langle x_Cx_D \rangle=&\eta T_{AC}V_{x_{A_0}}+\eta T_{AC}\xi_{AC}+N_0+V_{el},\\
\langle p^2_D\rangle-\langle p_Cp_D \rangle=&\eta T_{BC}V_{x_{B_0}}+\eta T_{BC}\xi_{BC}+N_0+V_{el}.
\end{split}
\end{equation}
In particular, we here assume that these detectors located in Charlie's side have identical detection efficiency $\eta$ and electronic noise $V_{el}$. Moreover, in the evaluation of the secret key rate, the above parameters $V_{x_{A_0}}$, $V_{x_{B_0}}$, $\xi_{AC}$, $\xi_{BC}$ and $V_{el}$ must be calculated in shot-noise unit, i.e., $V_AN_0$, $V_BN_0$, $\varepsilon_{AC}N_0$, $\varepsilon_{BC}N_0$ and $\nu_{el}N_0$, respectively.

By using Eqs. (\ref{eq5}) and (\ref{eq6}), the channel parameters $T_{AC}$, $T_{BC}$, $\varepsilon_{AC}$ and $\varepsilon_{BC}$ can be estimated by
\begin{equation}
\label{eq7}
\begin{split}
T_{AC}=&\frac{2\langle x_{A_0}x_C\rangle^2}{\eta \langle x^2_{A_0}\rangle^2},\\
T_{BC}=&\frac{2\langle p_{B_0}p_D\rangle^2}{\eta \langle p^2_{B_0}\rangle^2},\\
\varepsilon_{AC}=&\frac{\langle x^2_C\rangle + \langle x_Cx_D\rangle-N_0-\nu_{el}N_0}{2(\langle x_{A_0}x_C\rangle/\langle x^2_{A_0}\rangle)^2N_0}-\frac{\langle x^2_{A_0}\rangle}{N_0},\\
\varepsilon_{BC}=&\frac{\langle p^2_D\rangle - \langle p_Cp_D\rangle-N_0-\nu_{el}N_0}{2(\langle p_{B_0}p_D\rangle/\langle p^2_{B_0}\rangle)^2N_0}-\frac{\langle p^2_{B_0}\rangle}{N_0}.
\end{split}
\end{equation}
It is important to note that the quadrature variable $p_C$ of mode C or quadrature variable $x_D$ of mode D needs also to be measured for estimating these channel parameters. Therefore, we should use a heterodyne detector to replace one of the two homodyne detectors. More importantly, the above investigations are based on the fact that the CV-MDI-QKD can remove all side-channel attacks on detectors. Therefore, we consider that the precision of the estimated parameters are not limited by the detectors in Charlie's side.

However, the analyses of Sec. \ref{sec2} indicate that the estimated values may be affected by the laser seeding attack. In order to clearly show this influence, we here consider the worst case, which is that the two light source modules of the system is simultaneously attacked. For simplicity, we assume that the attack power is the same (i.e., ${g_1}={g_2}=g$) and do not consider the other attack situations. Therefore, these quadrature variable $x_{A_0}$ ($p_{A_0}$), $x_{B_0}$ ($p_{B_0}$), $x_C$ ($p_C$) and $x_D$ ($p_D$) will become
\begin{equation}
\label{eq8}
\begin{split}
x^\prime_{A_0}=&\sqrt{g}x_{A_0},p^\prime_{A_0}=\sqrt{g}p_{A_0},\\
x^\prime_{B_0}=&\sqrt{g}x_{B_0},p^\prime_{B_0}=\sqrt{g}p_{B_0},\\
x^\prime_{C}=&\sqrt{g}x_{C},p^\prime_{C}=\sqrt{g}p_{C},\\
x^\prime_{D}=&\sqrt{g}x_{D},p^\prime_{D}=\sqrt{g}p_{D}.
\end{split}
\end{equation}

In a practical CV-MDI-QKD system, if Alice and Bob are not aware of the laser seeding attack, they still use $x_{A_0}$ and $p_{B_0}$ to estimate the channel parameters. Therefore, these relations in Eq. (\ref{eq7}) kept by Alice, Bob and Charlie become
\begin{equation}
\label{eq9}
\begin{split}
T^\prime_{AC}=&\frac{2\langle x_{A_0}x^\prime_C\rangle^2}{\eta \langle x^2_{A_0}\rangle^2}=gT_{AC},\\
T^\prime_{BC}=&\frac{2\langle p_{B_0}p^\prime_D\rangle^2}{\eta \langle p^2_{B_0}\rangle^2}=gT_{BC},\\
\varepsilon^\prime_{AC}=&\frac{\langle (x^\prime_C)^2\rangle + \langle x^\prime_Cx^\prime_D\rangle-N_0-\nu_{el}N_0}{2(\langle x_{A_0}x^\prime_C\rangle/\langle x^2_{A_0}\rangle)^2N_0}-\frac{\langle x^2_{A_0}\rangle}{N_0}=\frac{\varepsilon_{AC}}{g},\\
\varepsilon^\prime_{BC}=&\frac{\langle (p^\prime_D)^2\rangle - \langle p^\prime_Cp^\prime_D\rangle-N_0-\nu_{el}N_0}{2(\langle p_{B_0}p^\prime_D\rangle/\langle p^2_{B_0}\rangle)^2N_0}-\frac{\langle p^2_{B_0}\rangle}{N_0}=\frac{\varepsilon_{BC}}{g}.
\end{split}
\end{equation}

There are some obvious deviations caused by the laser seeding attack between Eq. (\ref{eq7}) and Eq. (\ref{eq9}). It is obvious that these channel excess noises are underestimated under the laser seeding attack. Therefore, Eve may perform a classical intercept-resend attack to collect key information without trace under the shield of the laser seeding attack, which illustrates that a loophole will occurs in a practical CV-MDI-QKD system. In order to clearly show the loophole, we cite a specific partial intercept-resend (PIR) attack to analyze the security of a practical CV-MDI-QKD system in presence of the laser seeding attack in the next section.
\subsubsection{A quantitative example}
The intercept-resend attack plays an important role as one part of most quantum hacking strategies. In the quantum hacking scheme based on the laser seeding attack, Eve may also exploit the classical intercept-resend attack to collect key information. Therefore, we first investigate the PIR attack between Alice and Charlie under the laser seeding attack. In the PIR attack, the probability distribution of quadrature variable of mode $A^\prime$ in Charlie's apparatus is weighted sum of two Gaussian distributions, i.e., the distribution of the intercepted resend data with a weight of $u$ and the distribution of the transmitted data with a weight of $1-u$ \cite{Jouguet2013Preventing,lodewyck2007experimental}. Further, the extra excess noise caused by Eve in the implementation of the PIR attack can be expressed by $2uN_0$. In principle, the total excess noise estimated by Alice and Charlie under the PIR attack can be represented as
\begin{equation}
\label{eq10}
\xi_{PIR,AC}=\xi_{t,AC}+2uN_0,
\end{equation}
where $\xi_{t,AC}=\varepsilon_{t,AC} N_0$ is the technical excess noise. Expressed in shot-noise, the estimated excess noise $\xi_{PIR,AC}$ can be computed as
\begin{equation}
\label{eq11}
\varepsilon_{PIR,AC}=\varepsilon_{t,AC}+2u.
\end{equation}
With loss of generality, we assign 0.1 to $u$. Correspondingly, the excess noise $\varepsilon_{PIR,AC}$ estimated by Alice and Charlie can become $\varepsilon_{t,AC}+0.2$. In this case, the estimated excess noise under the laser seeding attack should be rewritten as
\begin{equation}
\label{eq12}
\varepsilon^\prime_{PIR,AC}=\frac{\varepsilon_{t,AC}+0.2}{g}.
\end{equation}

In the practical implementation of a CV-MDI-QKD system, we assume that the technical excess noise $\varepsilon_{t,AC}=0.1$. Therefore, when Eve performs the PIR attack, the estimated total excess noise under the laser seeding attack can be calculated as $\varepsilon^\prime_{PIR,AC}=\frac{0.3}{g}$. Before the execution of the laser seeding attack, the noise value is $0.3$, which obviously exceeds the ideal value. Accordingly, the process of key distribution is interrupted to guarantee the security of the system. However, we find that the estimated total excess noise can be reduced by Eve with the help of the laser seeding attack. When $g=3$, the estimated total excess noise $\varepsilon^\prime_{PIR,AC}=0.1$, i.e., the ideal noise value without attack. It has been experimentally demonstrated that $g$ can equal 3 under the control of Eve \cite{huang2019laser}. The result indicates that Eve can perform the laser seeding attack to make the PIR attack hidden. In particular, Eve can perform a full intercept-resend (FIR) attack in the case of $u=1$. Although the FIR attack is the most powerful, it can also be hidden when Eve makes $g$ exceeds 21. These analysis results fully demonstrate that the extra excess noise induced by the intercept-resend attack can be completely concealed by Eve through the laser seeding attack. Similarly, the excess noise induced by the intercept-resend attack between Bob and Charlie can also be concealed by Eve with the help of the laser seeding attack. Therefore, the laser seeding attack will open a loophole for Eve to successfully hide her attacks, which seriously destroys the security of the practical CV-MDI-QKD system.
\subsubsection{Secret key rate under the laser seeding attack}
In this section, we mainly focus on the secret key rate of a CV-MDI-QKD scheme under one-mode collective Gaussian attack, where Bob performs reverse reconciliation. We here point out that this one-mode attack is not the optimal strategy. In particular, the two-mode attack is demonstrated to be the optimal attack \cite{pirandola2015high-rate}. More concretely, Eve performs correlated two-mode coherent Gaussian attack on two quantum channels by employing their interactions. However, in a practical CV-MDI-QKD system, the correlation of the two quantum channels can become very weak when they come from different directions. Therefore, the quantum channel of CV-MDI-QKD can be reduced to one-mode channel in this context. Here, Eve can efficiently perform the one-mode attack.

It has been demonstrated that the CV-MDI-QKD schemes is equivalent to the one-way CVQKD protocols using coherent states and heterodyne detection when the preparation of Bob's EPR states and the displacement operation of Bob are assumed to be untrusted \cite{li2014continuous-variable}. Therefore, the calculation process of secret key rate of CV-MDI-QKD protocols is the same with the one-way GMCS CVQKD. In the following analysis, we assume that the heterodyne detection is perfect and do not consider the finite-size effect, which does not affect our analysis results. Here, the shannon mutual information between Alice and Bob becomes \cite{fossier2009improvement}
\begin{equation}
\label{eq13}
I^{het}_{AB}=2\times \frac{1}{2}\log_2\frac{V^{het}_B}{V^{het}_{B|A}}=\log_2\frac{V_A+1+\chi_{\text{line},m}}{1+\chi_{\text{line},m}},
\end{equation}
where $V^{het}_B=\eta T_m(V_A+1+\chi_{\text{line},m})/2$, $V^{het}_{B|A}=\eta T_m (1+\chi_{\text{line},m})/2$ and $\chi_{\text{line},m}=1/T_m-1+\varepsilon_m$. Furthermore, the vital covariance matrix ${\Gamma^m _{AB}}$ between Alice and Bob can be expressed as
\begin{equation}
\label{eq14}
\begin{split}
&{\Gamma^m _{AB}}=\\
&\left[ {\begin{array}{*{20}{c}}
{\left( {{V_A} + 1} \right){\mathbb {I}}}&{\sqrt {T_m\left[ {{{\left( {{V_A} + 1} \right)}^2} - 1} \right]} {\sigma _Z}}\\
{\sqrt {T_m\left[ {{{\left( {{V_A} + 1} \right)}^2} - 1} \right]} {\sigma _Z}}&(T_m{V_A} + 1 + T_m\varepsilon_m)\mathbb {I}
\end{array}} \right],
\end{split}
\end{equation}
where
\begin{equation}
\label{eq15}
\begin{split}
T_m =& \frac{T_{AC}}{2}{k^2},\\
\varepsilon_m =& 1 + \frac{1}{T_{AC}}[2 + T_{BC}(\varepsilon _{BC} - 2) + T_{AC}(\varepsilon _{BC} - 1)]\\
 &+ \frac{1}{T_{AC}}\left(\frac{\sqrt 2 }{k}\sqrt {V_B}  - \sqrt {T_{BC}} \sqrt {{V_B} + 2}\right)^2.
\end{split}
\end{equation}
In order to minimize $\varepsilon_m$, we adopt $k = \sqrt {\frac{{2{V_B}}}{{{T_{BC}}\left( {{V_B} + 2} \right)}}}$; then
\begin{equation}
\label{eq16}
\varepsilon_m = \frac{{{T_{BC}}}}{{{T_{AC}}}}\left( {{\varepsilon _{BC}} - 2} \right) + {\varepsilon _{AC}} + \frac{2}{{{T_{AC}}}}.
\end{equation}

According to Refs. \cite{fossier2009improvement,leverrier2010finite}, the Holevo bound can be obtained as
\begin{equation}
\label{eq17}
\chi_{BE}=G(\frac{\lambda_{m,1}-1}{2})+G(\frac{\lambda_{m,2}-1}{2})-G(\frac{\lambda_{m,3}-1}{2}).
\end{equation}
Here,
\begin{equation}
\label{eq18}
\begin{split}
\lambda^2_{m,1,2}&=\frac{1}{2}(A_m\pm\sqrt{A^2_m-4B_m}),\\
\lambda_{m,3}&=\frac{({T_m}{\varepsilon_m}+2)({V_A}+1)-{T_m}{V_A}}{{T_m}({\varepsilon_m}+{V_A})+2},
\end{split}
\end{equation}
where
\begin{equation}
\label{eq19}
\begin{split}
A_m=&(V_A+1)^2-2T_m(V^2_A+2V_A)\\
&+({T_m}{V_A}+T_m\varepsilon_m+1)^2,\\
B_m=&[({T_m}{\varepsilon_m}+1)(V_A+1)-{T_m}{V_A}]^2.
\end{split}
\end{equation}

Eventually, the secret key rate against collective attacks for the CV-MDI-QKD schemes is calculated as
\begin{equation}
\label{eq20}
K_m=\beta I^{het}_{AB}-\chi_{BE}.
\end{equation}

The analysis indicates that the secret key rate of the system can be expressed as $K_m=K(V_A,V_B,T_m,\varepsilon_m)$. When Alice and Bob are not aware of the laser seeding attack, the evaluative secret key rate is expressed as $K_{m,e}=K(V_A,V_B,T_m,\varepsilon^\prime_m)$. However, the practical secret key rate of the system should be computed as $K_{m,p}=K(V^\prime_A,V^\prime_B,T^\prime_m,\varepsilon_m)$. Here,
\begin{equation}
\label{eq21}
\begin{split}
V^\prime_A&=gV_A,V^\prime_B=gV_B,\\
T^\prime_m&=\frac{gT_{AC}V_B}{T_{BC}(gV_B+2)},\\
\varepsilon^\prime_m&=\frac{T_{BC}}{T_{AC}}(\frac{\varepsilon_{BC}}{g}-2)+\frac{\varepsilon_{AC}}{g}+\frac{2}{gT_{AC}}.
\end{split}
\end{equation}
\begin{figure}[!h]\center
\centering
\resizebox{8.5cm}{!}{
\includegraphics{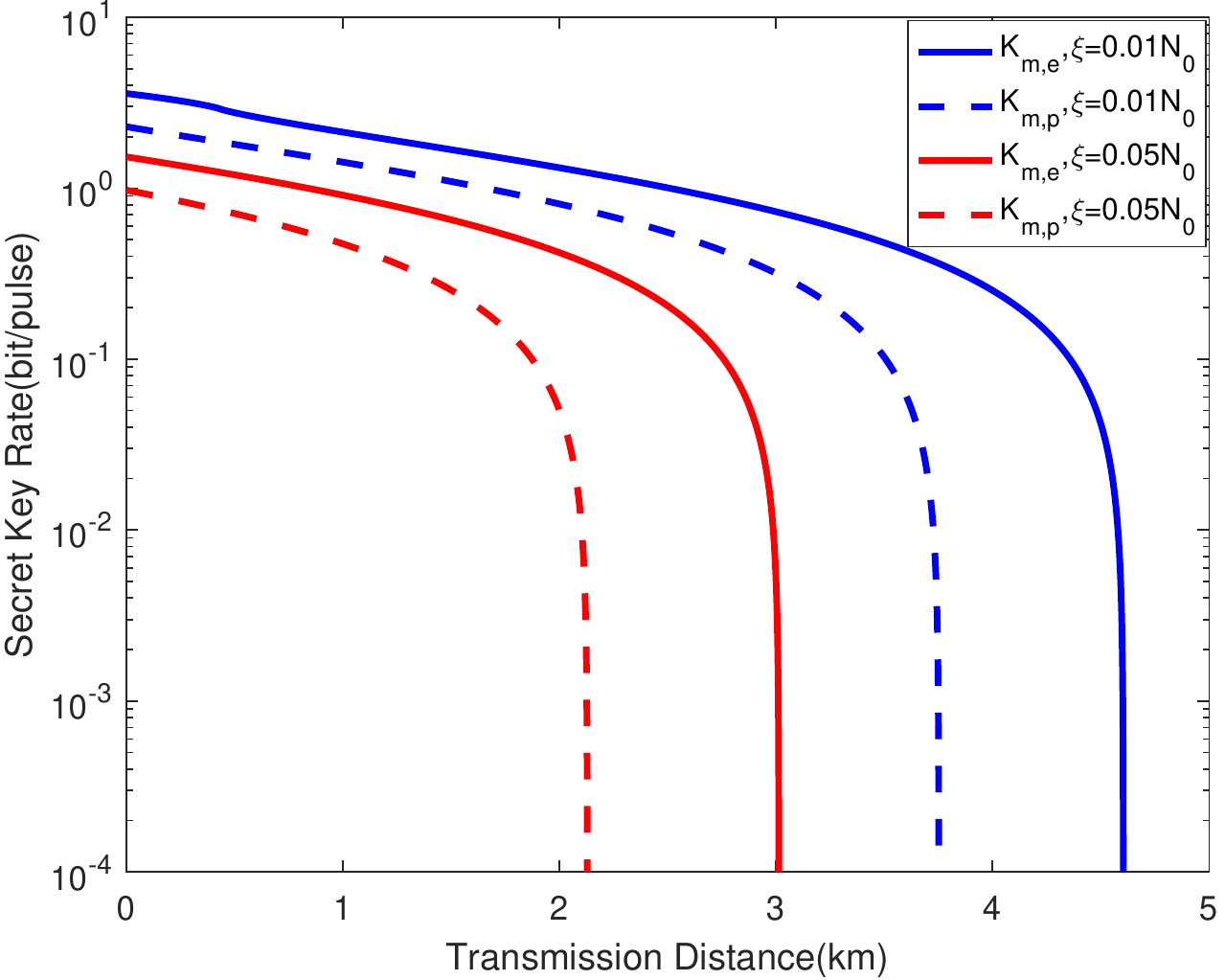}}
\caption{Secret key rate as a function of the transmission distance from Alice to Bob for different excess noise in the symmetric case, where $L_{AC}=L_{BC}$. Solid curves from top to bottom represent the relations between the evaluated secret key rate $K_{m,e}$ and the transmission distance when $\varepsilon_{AC}=\varepsilon_{BC}=0.01, 0.05$. Dotted curves show the corresponding practical secret key rate $K_{m,p}$ versus transmission distance. The fiber loss is 0.2 dB/km.}
\label{FIG5}
\end{figure}

Next, we simulate the secret key rate versus transmission distance in the symmetric and extreme asymmetric case when Eve performs the laser seeding attack in the two light source modules of a practical CV-MDI-QKD system. Fig. \ref{FIG5} shows the secret key rate versus transmission distance in the symmetric case under different excess noise environment when the attack power $g=1.02$. In the simulation analysis, the involved parameters are fixed as follows. ${V_A}={V_B}=40$, $\eta=0.6$, $\varepsilon_{AC}=\varepsilon_{BC}=0.01, 0.05$, $\beta=95\%$, respectively. We observe that there is an obvious gap between the secret key rate estimated by Alice and Bob and the practical secret key rate. The results indicate that Eve can perform the intercept-resend attack without trace to steal key information in a practical CV-MDI-QKD system.
\begin{figure}[!h]\center
\centering
\resizebox{8.5cm}{!}{
\includegraphics{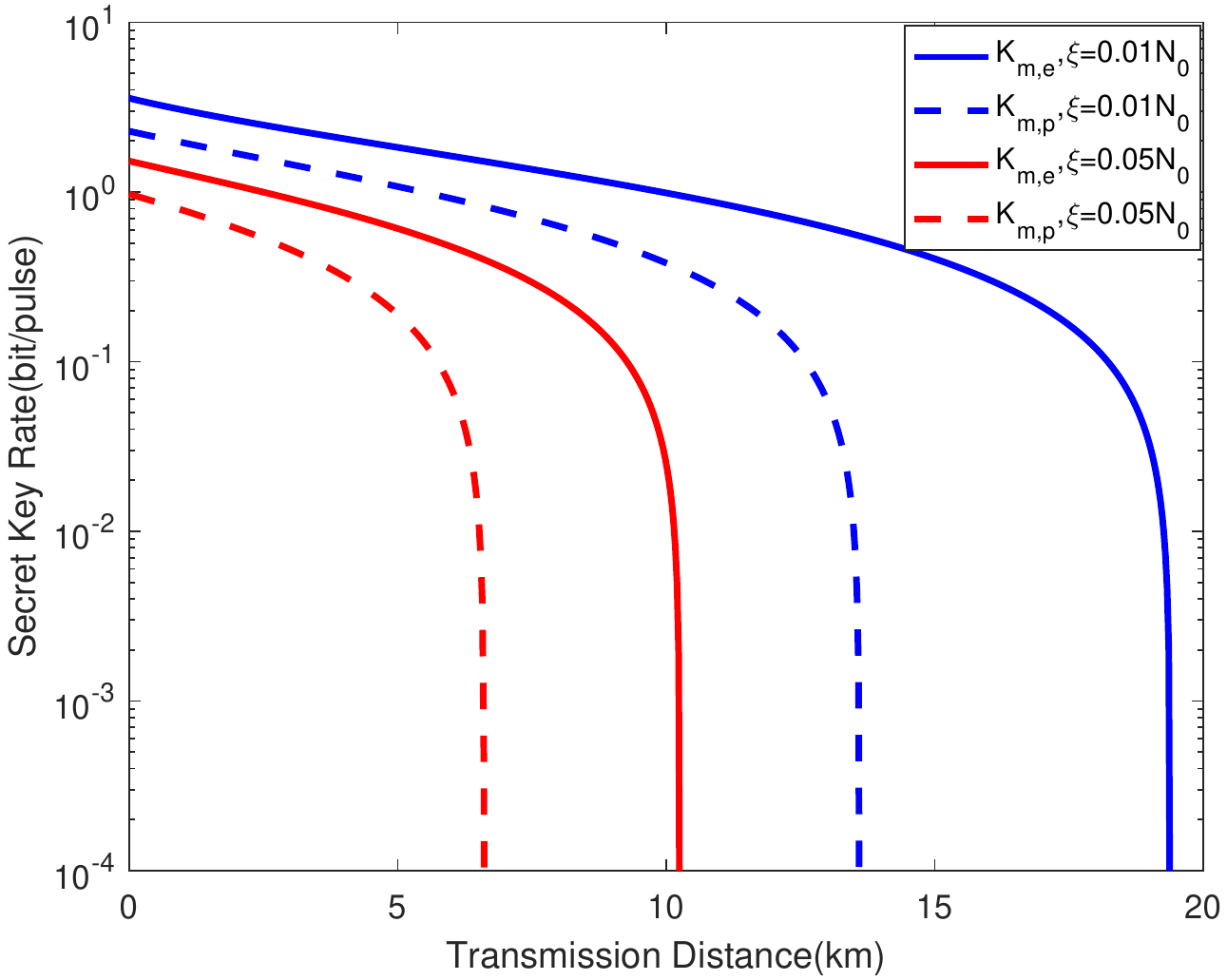}}
\caption{Secret key rate vs the transmission distance from Alice to Bob for different excess noise environments in the extreme asymmetric case, where $L_{BC}=0$. Solid curves from top to bottom represent the relations between the evaluated secret key rate $K_{m,e}$ and the transmission distance when $\varepsilon_{AC}=\varepsilon_{BC}=0.01, 0.05$. Dotted curves show the corresponding practical secret key rate $K_{m,p}$ versus transmission distance.}
\label{FIG6}
\end{figure}

Fig. \ref{FIG6} reveals the secret key rate of the system as a function of the transmission distance from Alice to Bob in the extreme asymmetric case when the excess noise $\varepsilon_{AC}$ and $\varepsilon_{BC}$ are assumed to be 0.01 or 0.05. Here, the attack power $g=1.02$, and the other parameters for the simulation are fixed values that has been confirmed in the analysis of the symmetric case. It is obvious that the secret key rate calculated by Alice and Bob is overestimated compared with the practical secret key rate. In particular, we find that the CV-MDI-QKD systems are more sensitive to the laser seeding attack than the one-way GMCS CVQKD system. A slight attack power can have a major impact on the evaluative value of secret key rate of the CV-MDI-QKD systems, especially in the extreme asymmetric case.

The above investigations demonstrate that the laser seeding attack opens a security loophole for Eve to obtain information about secret key without trace in a practical CV-MDI-QKD system, which seriously destroys the practical security of the system.
\section{Countermeasure against the laser seeding attack}\label{sec4}
The above investigations show that the laser seeding attack affects the parameter estimation and the evaluative secret key rate. To resist this attack, we can exploit an appropriate isolator to prevent the injected light. However, it is important to note that Eve might reduce the performance of the isolator by laser damage attack. Therefore, we here propose a real-time monitoring scheme for the intensity of output optical signal in light source module to prevent the incorrect estimation of channel parameters. According to the analysis of Section. \ref{sec2}, we find that the intensity of the LO signal can simultaneously change under the effects of the laser seeding attack. Therefore, the attack can be directly found by monitoring the intensity of the LO signal in real time before attenuation.

\begin{figure}[!h]\center
\centering
\resizebox{8.5cm}{!}{
\includegraphics{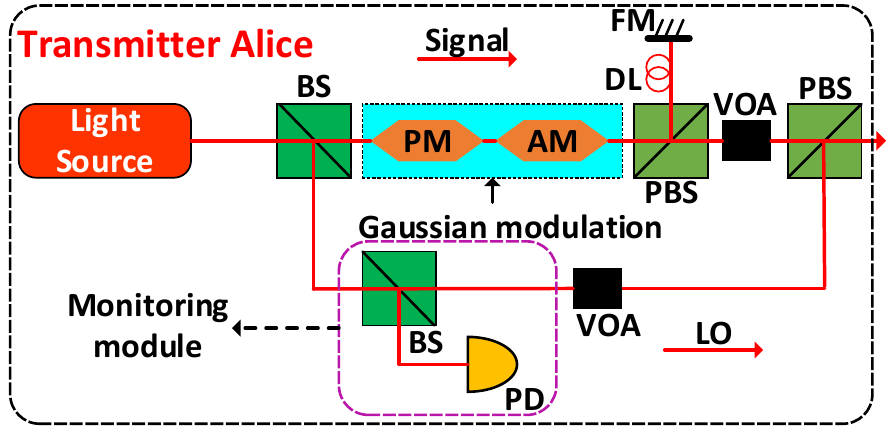}}
\caption{The structure of the real-time monitoring scheme against the laser seeding attack in Alice's apparatus for the one-way CVQKD stsyems. AM, amplitude modulator; PM, phase modulator; BS, beam splitter; PBS, polarizing beam splitter; FM, faraday mirror; DL, delay line; LO, local oscillator; VOA, variable optical attenuator; PD, photodiode.}
\label{FIG7}
\end{figure}
Fig. \ref{FIG7} shows the countermeasure against the laser seeding attack for the one-way CVQKD systems in Alice's apparatus. Specifically, Alice first splits a fraction of the undiminished LO signal to measure its intensity by using a photodiode. Here, the practical value of the intensity of the undiminished LO signal is $I_p$, which can also be automatically predicted by using machine learning \cite{liu2018}. Then, Alice can calculate the value of $g$ by comparing the difference between the measured value and the preset value $I_0$ of the separated LO signal, i.e., $g=\frac{I_p}{I_o}=\frac{V^\prime_{A_0}}{V_{A_0}}$. Eventually, according to Ref. \cite{zheng2019practical}, Alice and Bob can precisely evaluate the secret key rate of the system, i.e., $K_m=K(gV_{A_0}, \frac{T^\prime}{g}, g\varepsilon^\prime, \nu_{el})=K_p$. These analyses demonstrate that the real-time monitoring scheme can help Alice and Bob to precisely estimate the channel parameters of the system. Finally, the secret key rate of the CVQKD systems under the laser seeding attack can be precisely evaluated with the help of this scheme. The accurate evaluation of secret key rate can effectively close this security loophole. In particular, in a LLO CVQKD system, Alice can split a fraction of undiminished reference signal to monitor its intensity in real-time to close this loophole. In addition, the loss of the LO signal in the above scheme can be completely compensated by properly adjusting the preset value of the attenuation level of the VOA in the light path of LO.

It is important to note that this real-time monitoring scheme can equally remove the loophole induced by the laser seeding attack in a practical CV-MDI-QKD system. Here, the two light sources of the system should be simultaneously monitored by adding the monitoring module which is shown in Fig. \ref{FIG7}. More concretely, the monitoring scheme also makes the secret key rate evaluated precisely to resist the laser seeding attack. For example, in the case of the same attack power, the parameter $g$ can be acquired by $g={g_1}={g_2}=\frac{I_{p_1}}{I_0}=\frac{I_{p_2}}{I_0}$. Next, the secret key rate of the system can be evaluated as $K_{m,e}=K(gV_A, gV_B, T^\prime_m, \varepsilon_m)=K_{m,p}$. The analysis result indicates that the proposed real-time monitoring scheme can effectively resist the laser seeding attack to close this security loophole.
\section{CONCLUSION}\label{sec5}
In this work we have studied the security of several practical CVQKD systems under the laser seeding attack. More specifically, we have studied the standard one-way GMCS CVQKD protocols in reverse reconciliation, and GMCS CV-MDI-QKD schemes in the symmetric case and extreme asymmetric case. Here, we consider that Eve can carry out the laser seeding attack in the laser sources of the two kinds of CVQKD systems. We show that the intensity of the transmitted Gaussian-modulated quantum optical signals can become large with the increase of the intensity of optical signals prepared by the light source module under the attack.

For the practical one-way CVQKD systems, we observe that the effects of the laser seeding attack are similar with the influences of the reduced optical attenuation caused by laser damage attack. Therefore, the laser seeding attack opens a loophole for Eve in the system. We further show that the laser seeding attack makes the secret key rate of the system overestimated, which also demonstrates the attack can help Eve to hide herself. In particular, Eve can obtain more key information for the case of a larger channel excess noise in the same attack power. In order to close this loophole, we propose a real-time monitoring scheme for the intensity of the optical signal generated by the light source module by measuring the intensity of the LO signal before attenuation. This scheme can make Alice and Bob precisely evaluate the channel parameters to accurately analyse the performance of the system.

Apart from this, we mainly investigate the laser seeding attack for the effects of the security of a practical CV-MDI-QKD system. We find that these channel excess noises of the system are underestimated under the laser seeding attack, which indicates that the attack can open a security loophole for Eve to successfully perform an intercept-resend attack. Although the CV-MDI-QKD system can remove all side channels from the measurement unit, Eve can also successfully perform the laser seeding attack in the two light source modules of the system to steal key information without being detected. We also find that the CV-MDI-QKD schemes are more sensitive to the laser seeding attack compared with the one-way CVQKD protocols. It is notable that the proposed real-time monitoring scheme can also close this loophole in a practical CV-MDI-QKD system.
\section*{ACKNOWLEDGMENTS}
This work was supported by the National key research and development program (Grant No. 2016YFA0302600), the National Natural Science Foundation of China (Grants No. 61671287, 61631014, 61332019, 61632021), and the Natural Science Basic Research Plan in Shaanxi Province of China (Grant No. 2019JM-591).

\end{document}